\documentstyle[12pt,epsfig]{article}
\date{}
\title{\normalsize{\bf{ON VOLUME  REFLECTION OF ULTRARELATIVISTIC  PARTICLES IN
SINGLE  CRYSTALS
}}}

\author{ V.A. Maisheev$^1$\\
 $^{1}$\it{ Institute for High Energy Physics, 142281, Protvino, Russia }}

\begin{document}
\maketitle

\begin{abstract}
The analytical description of the volume reflection of the charged
ultrarelativistic particles in bent single crystals is considered.
The relation describing the angle of volume reflection 
as  a function of the transversal energy is obtained.  
The different angle distributions of the scattered protons
in the single crystals are found. Results of calculations for
400 GeV protons scattered by the silicon single crystal
are presented. 
\end{abstract}

\section{Introduction}
The volume reflection of the charged particles in single crystals represents the  coherent scattering of these particles
by planar or axial electric fields of  bent crystallographic structures. For the first time, this effect was predicted in Refs.\cite{TV,TV1}. 
The angle of the volume reflection is small enough (in order of the critical angle of the channeling).
Because of this, it is thought that this process did not show itself in experimental measurements.
However, in the recent reports  \cite{Fli,Mo} the problems were discussed, which require the existing of volume reflection effect
 for explanation of  proton collimation  measurements.  Besides, the preliminary results of the experimental
observation of the volume reflection were published \cite{Iv, WS1}. 

The recent meeting in CERN \cite{CE} which was devoted to problems of utilization of interactions of proton beams
 with single crystals attract considerable interest to the phenomenon of volume reflection due to new
possibilities such as the collimation of collider beams \cite{WS, ADK, RWA}, creation of devises which
 can multiple the effect \cite{VG} and so on.
In the immediate future the experimental study of the volume reflection effect  are scheduled 
on extracted proton beams in CERN \cite{MF} and IHEP\cite{Che}.  

By now the main results of the theoretical investigation of the volume reflection are mainly  based  on  the Monte Carlo calculations.
No doubt, this is very powerful method, however,  the analytical solution of the problem is very useful for understanding
of the process and  its different characteristics.. In this paper  we consider such description of  the volume reflection effect . 

\section{ Motion in bent single crystals}
The motion of the ultrarelativistic particles in the bent single crystals one can
describe with the help of the following equations\cite{Ts,XA} :
\begin{equation}
  E_0 \beta^2 v_r^2/(2c^2) +U(r)+ E_0 \beta^2 (r-R)/r= E=const
\end{equation}
\begin{equation}
dy/dt=v_y=const
\end{equation}
\begin{equation}
v_z = rd\phi/dt \approx  c(1-(1/\gamma^2+v^2_r + v_y^2)/2)
\end{equation}
These equations take place for cylindrical coordinate system ($r, \phi, y$). 
Here $v_r$ is the component of particle velocity in the radial direction, $v_y$ is the component
of the velocity along $y$-axis and $v_z$ is the tangential component of the velocity,
$R$ is the radius of bending of the single crystal, $E_0$  and $\gamma$ are the particle energy and
its Lorentz factor, $E$ is the constant value of the radial energy, $U(r)$ is the one dimensional potential
of the single crystal, $c$ is the velocity of light and $\beta$ is the ratio of the particle velocity
to  velocity of light. In this paper we consider the planar case when the scattering 
is due to interaction particles with the set of the crystallographic planes located normally to ($r, \phi$)-plane . 
On the practice it means that $v_y/c \gg \theta_{ac}$ but $v_y/c \gg 1$ for ultrarelativistic particles,
where $\theta_{ac}$ is the critical angle of the axial channeling.

Eq.(1) one can transform in the following
form:
\begin{equation}
E_0 \beta^2 v_x^2/(2c^2) + U(x) +E_0\beta^2 x/R=E
\end{equation}
where $x$ is the local Cartesian coordinate  which connected with the cylindrical coordinate $r$
through the relation  $x= r-R$ and $v_x=v_r$.  We also change $r$-value in the denominator of Eq.(1) on R.
For real experimental situation it brings negligible mistake (in order of x/R).
In Eq.(4)  E-value have a sense of the transversal energy.

The considered here equations describe the three dimensional motion of the particle in the
bent single crystal in the cylindrical coordinate system.  Let us introduce the Cartesian coordinate
system in which the xy-plane is coincident with the front edge of  a single crystal. Now
we can calculate  x-component of the velocity in this system:
\begin{equation}
V_x= v_x \cos\phi +v_z\sin \phi \approx  v_x+v_z \phi
\end{equation}

Fig.1 illustrates the geometry  of proton beam volume reflection and different  coordinate systems
which will be used in our consideration.

\section{Angle of volume reflection}
One can see that Eq.(4) describes one dimensional motion  in the effective potential 
$U_e(x)= U(x)+E_0\beta^2x/R$. There are two different  kinds of the infinite motion in the effective potential.
One  case corresponds to motion when particle after its entrance in single crystal move
in the direction of the increasing of potential ( in positive direction of x-axis). 
In this case the velocity $v_x$ decrease and in some point $x_c$ become equal to zero
and then particle begin to move in the opposite direction. It  corresponds to case
when effect of volume reflection is possible. 
In the second case the particle after entrance in single crystal move in the 
negative direction of the x-axis and volume reflection is impossible.  
Obviously, that the belonging  of a  trajectory to some kind of events is determined by the
initial conditions of particles at entrance in the single crystal: the first case takes place
at the negative entering angles $\theta$ relative to crystallographic planes.
We also will consider below case when the absolute value of  $\theta$ is large than
channeling angle $\theta_c$.  

The trajectory in x-direction  ($ x_0 \le x \le x_c$) up to point $x_c $ one can calculate as
\begin{equation}
t=\int_{x_0}^{x}{dx \over \sqrt{{2c^2\over E_0\beta^2} ( E-U(x)-E_0\beta^2 x /R)                      
}}\,,   
\end{equation}
where $t$ is the time and $t=0$ corresponds to coordinate $x_0$ at entrance of the single crystal.
We can rewrite this equation in the following equivalent form:
\begin{equation}
t=\int_{x_0}^{x}[{1 \over \sqrt{{2c^2\over E_0\beta^2} ( E-U(x)-E_0\beta^2 x /R)}} - F(x)]dx
+\int_{x_0}^{x} F(x)dx  
\end{equation}  
where the  function $ F(x)$ reads
\begin{equation}
F(x)={1 \over \sqrt{{2c^2\over E_0\beta^2} ( E-U(x_c)-E_0\beta^2 x /R)}}
\end{equation}
and the critical point $x_c$ is determined by the equation:
\begin{equation}
 E-U(x_c)-E_0\beta^2 x_c /R =0
\end{equation}
Note that equation $t = \int_{x_0}^{x} F(x)dx $ describes the one dimension motion of a particle,
which do not interact with the electric field of the single crystal. We name this case as the neutral motion.

Taking the integral $ \int_{x_0}^{x} F(x)dx $ we get 
\begin{equation}
t=T(x_0,x)  +R/c^2(\tilde{v}(x_0)-\tilde{v}(x))
\end{equation}
where
\begin{equation}
T(x_0,x)=\int_{x_0}^{x}[{1 \over \sqrt{{2c^2\over E_0\beta^2} ( E-U(x)-E_0\beta^2 x /R)}} - F(x)]dx
\end{equation}
and $\tilde{v}(x)$ is the velocity of the neutral motion in the point $x$ of its trajectory:
\begin{equation}
\tilde{v}(x)=\sqrt{ 2c^2/(E_0 \beta^2)( E-U(x_c)-E_0\beta^2 x /R)}
\end{equation}  

The angle $\phi$ in Eq.(5) is approximately equal to $V_zt/R \approx Vt/R \approx ct/R$
For our aims the small difference between $V_z, V, c$ is of no significance. We will
use the relation: $\phi=ct/R$.  Hence,  Eq.(5) have the following form:
\begin{equation}
V_x=v_x+ v_z ct /R \approx v_x+  c^2t/R
\end{equation}
Taking t from Eq.(10) one can get
\begin{equation}
V_x =v_x + T(x_0,x)c^2/R+ \tilde{v}(x_0)-\tilde{v}(x)
\end{equation} 
Now we assume that  the distance $d_{vr}= |x_c-x_0|$ is  large enough. According to estimation in Ref.\cite{TV1}
$d_{vr} \gg \theta_{c}  R$. Hence, $V_x\approx \tilde{v}(x_0) $  ($ v_x \approx  \tilde{v}(x), \, T(x_0,x) \approx0$)
 when  the particle move far from the critical point  (near $x_0$ point).
It easy to see that
in the critical point $v_x=0$, $\tilde{v}(x)=0$ and $V_x= \tilde{v}(x_0) +T(x_0,x_c)c^2/R$. 
Using the obvious symmetrical properties of the one dimensional motion we can get
that for distant enough from critical point  values $ x>x_c$  $V_x= \tilde{v}(x_0) +2T(x_0, x_c)c^2/R$.  
From here the angle of volume reflection $\alpha = 2T(x_0, x_c) c /R$  or its explicit 
value have the following form: 
\begin{equation}
\alpha ={2c \over R} \int_{x_0}^{x_c}[{1 \over \sqrt{{2c^2\over E_0\beta^2} ( E-U(x)-E_0\beta^2 x /R)}} - F(x)]dx
\end{equation}

It is easy to understand the meaning of the F-function introduced in Eq.(8) .  This function allows one
to eliminate the influence of the geometrical factors on the particle motion in the cylindrical coordinate
system.  The visual picture of volume reflection may be also observed in cylindrical coordinates $r$ and $\phi$.
Really the trajectories of the charged and neutral particles have the same critical radial coordinate $r_c$
but the different values of $\phi_c$. The difference of $\phi_c$-coordinates determines
the angle of volume reflection.

\section{Periodicity of volume reflection angle}
Considering  the volume reflection angle as a function of the transversal energy $E$ one can see
that this function is periodic. The period is equal to 
\begin{equation}
\delta E =E_0\beta^2d/R.
\end{equation}
 This fact one can get
by the substitution $ x= x+d$ (d is the interplanar distance) and taking into account the
periodicity of potential $U(x)$.

The transversal energy $E$ have the following representation:
\begin{equation}
E= E_0 \beta^2 \theta^2/2 + U(x_0)+ E_0\beta^2 x_0 /R
\end{equation}
The replacement $ x_0$ on $ x_0+ d$ corresponds to new energy  $E+\delta E$. It means
that  the angle $\alpha$ is periodic function of the initial coordinate $x_0$. 

The volume reflection angle is also a periodic function of the square of initial angle $\theta$.
In this case the period is  
\begin{equation}
\delta \theta^2  = 2d/R  
\end{equation}
It should be noted that periodicity of  volume reflection angle relative to above pointed parameters violated if the condition
$d_{vr} \gg \theta_c R$  is not satisfied  or, in other words,  at transformation from one energy to
another the distance $|x_0- x_c|$ should be  large enough.

\section{Particle distributions at volume reflection}

In calculations  we  will assume that the distribution of the particle beam over the initial coordinate $x_0$
is uniform  (in the limits of one period). If the function $\alpha(x_0)$  ($0 \le x_0 \le d$) is known
we can find the distribution of particles over the angle of volume reflection with the help
of the following relation:
\begin{equation}
{dN \over d\alpha}(\alpha)=\sum_i {1 \over d}|{d\alpha \over dx_0}|^{-1}
\end{equation}
where the sum over i means that the sum should be taken over every domain of the function single-valuedness
and the derivative $d\alpha / dx_0$ should be calculated for values $x_{0i}$ which
satisfy to equation $\alpha(x_{0i})=\hat{\alpha}$  in which $\hat{\alpha}$ is the current
(computed) value of the volume reflection angle.  Note Eq.(19) was obtained at the condition that the initial
angle $\theta$ is fixed. Besides, in this paper we use the normalized to 1 the distribution functions.

In actual practice,  the particle beam have some distribution over the initial angles.
According to our consideration the distribution function over $\alpha$ dependent on
the initial angle value. On the other hand,  the volume reflection angle is the periodic 
function of $\theta^2$-value. At large enough $\theta$-angles and small range of variation of
this angle  the angle $\alpha$ one can consider as periodic function of the initial
angle $\theta$ with the period:
\begin{equation}
\delta \theta={d \over \theta R}
\end{equation}
It easy to see that $\delta \theta$ is very small value. Simple estimations give 
values of $\delta \theta \sim  10^{-6}-10^{-8}$ radian for radius in the range 1- 100 m,
correspondingly,  for initial angle about $10^{-4}$ radian and $d$ about some angstroms.
The real particle beams have significantly more a broad angle distribution.   
Assuming the uniform distribution of the initial angles within the angle period $\delta \theta$
(at the assumption that divergence of beam satisfied the condition $ \Delta \theta \ll \theta$)
we get    the  distribution averaged over its period
\begin{equation}
<dN/d\alpha> = {1\over \delta \theta} \int_{0}^{\delta \theta} {dN \over d\alpha} d\theta 
\end{equation}
This equation is obtained for physically narrow particle beam. Under this term we understand the beam
with the small divergence of initial angles, which  in tens or more times exceed
the angle period in Eq.(20). The distribution of the beam with a valuable quantity $\Delta \theta$
(however, $\Delta \theta \ll \theta$) have the following form
\begin{equation}
 <{dN\over d\alpha} (\alpha) >_{\Delta \theta}= \int_{\theta_0 - \theta_1 }^{ \theta_2 - \theta_0}
 \rho_b (\xi, \theta_0) <{dN \over d\alpha}(\alpha -\xi)> d\xi 
\end{equation}
where $\rho_b (\xi, \theta_0)$ is the angle distribution of beam around the initial angle $\theta_0$,
$\theta_1$ and $\theta_2$ are minimal and  maximal values of the initial angles ($\Delta \theta= \theta_2-\theta_1$).
 In this equation the angle $\alpha$ is the planar angle relative to direction which is determined by $\theta_0$.

\section{Influence of multiple scattering}
The above-mentioned results of our consideration are appropriate for the ideal case only 
when the dissipative processes in the body of single crystal are absent.  
The Monte Carlo calculations  can give detail consideration of these processes.
Here we make attempt to take into account the influence of the multiple scattering
on the process of volume reflection in the frame of the analytical description.
However,  in this case some realistic assumptions is needed.
As  was said earlier the main contribution in integral describing the angle motion of particle
(see Eq. (15) ) give relatively not large a space area near the critical point $x_c$. Hence, we can assume
that we can neglect by the myltyple scattering in this area.

The process of multiple scattering on the particle path from $x_0$ till $\sim x_c$ bring
the variation of the transversal energy E.  Bearing  in  mind that the scattering can play
the important role in the small area near $x_c$-coordinate we suppose that this factor 
may be   taking into account by the corresponding changing initial data (see Eq.(17)).
We assume that the physically narrow beam have the initial angle $\theta$.
Then we can consider that  the initial distribution of particles is following:
\begin{equation}
<{dN\over d\alpha} (\alpha) >_{ms}= \int_{-\infty}^{\infty}
\rho(\varphi, \sigma) <{dN\over d\alpha}(\alpha - \varphi) > d\varphi
\end{equation}
where $\rho(\varphi)={1\over \sqrt{2\pi}\sigma} \exp -(\varphi^2/2\sigma^2)$ is angle distribution
of multiple scattering with the parameter $\sigma$  depending on the distance $\approx |x_c-x_0|/\theta$. 
We also expect that the redistribution of the initial coordinates is small and do not consider  this factor. 
Here it is important that the distribution of the particles within interplanar distance remains uniform.

Now we can obtain the angle distribution after single crystal for physically narrow beam:
\begin{equation}
<{dN\over d\alpha} (\alpha) >_{exit}= \int_{-\infty}^{\infty}
\rho(\varphi, \sigma_1) <{dN\over d\alpha}(\alpha - \varphi) > _{ms}d\varphi
\end{equation}
where $\sigma_1$ depends on the distance $\approx |x_e-x_c|/\theta$ ($x_e$ is the exit coordinate). 
Obviously the similar relations one can get for more complicated distribution of the initial angles.

\begin{table}[t]
\begin{center}
\begin{tabular*}{133 mm}[ ]{| r | l | l | l | l | }
\hline
$R, m$ & $l_s, cm$ & $<\theta> 10^5 , radian$ & $<\delta\theta> 10^6, radian$ & 
$Remark $ \\
\hline
25 &$0$ & $1.52 $ &$0.63  $ & Fig. 6, curve 1  \\
25 &$0.125$ & $1.52 $ & $4.2 $ & Fig.7a, curve 1 \\
25 &$ 0.25$ & $ 1.51 $ & $5.8 $ & Fig.7a, curve 2 \\
25 &$0.5$ & $1.45 $ & $7.7 $ &Fig.7a, curve 3 \\
\hline
10 &$ 0  $ & $1.39 $ &$1.5$ & Fig. 6, curve 2 \\
10 &$0.05 $ & $1.38 $ &$3.1 $ & Fig.7b, curve 1 \\
10 &$0.1$ &$1.38 $ &$4.1 $&Fig.7b, curve 2 \\
10 &$0.2$ & $1.38 $ &$5.5 $&Fig. 7b, curve 3\\
\hline
5 &$ 0$&$1.14  $ &$3.2 $&Fig.6, curve 3\\
5 &$ 0.025$&$1.14  $ &$3.7 $&Fig.7c  curve 1\\
5 &$ 0.05$&$1.14  $ &$4.1 $&Fig.7c, curve 2\\
5 &$ 0.1$&$1.14 $ &$4.9 $&Fig.7c, curve 3\\
\hline
\end{tabular*}
\caption{ {\bf{ The initial conditions for calculations of the entering angle distributions and
some results of these calculations
}}  
\newline
R is the bending radius, $ l_s$ is the total length of the scattering,
$<\theta>$ and $<\delta\theta>$ are the mean angle and standard deviation  of 
the distribution.
}
\end{center}
\end{table}

\section{Calculations}
In this section we demonstrate applications of obtained  relations for calculations of 
the specific cases volume reflection effect.
All our calculations will be done for energy of the proton beam equal to 400 GeV,
which was planned in CERN channeling experiment  
and for the (110) plane of the silicon single crystal. 

We start with detail consideration of calculation for proton beams, and then we also touch
the case of negative particles. 
\subsection{Atomic potential for calculations}
In contrast to channeling regime the above-barrier particle move throughout all the points of the potential
and its influence on the motion of particle arises near critical point $x_c$, where the particle is located
close to top of potential barrier for long time.   
Because of this,  we think that  the correct choice of the one dimensional atomic potential is
important problem.  

Taking into account aforesaid we choose the one dimensional atomic potential according to Ref.\cite{MV},
where potential was calculated on the basis of the measurements of atomic form factors 
with the use of  x-rays \cite{ CW}.
For this aim  we use the expansion of the potential in the Fourier series. 
In a similar manner we find the one dimensional electric field and electron density.
All these characteristics were calculated at room temperature. 
It should be noted that this potential was used for comparison of the theoretical
description and experimental results for coherent bremsstrahlung  of 10 GeV positrons 
for  the (110) plane of the silicon single crystal  \cite{NKB} and in this work was shown very good agreement between theory
and measurements. 

We also calculate the one dimensional potential and electric fields  using the  Moliere approximation of atomic form factors.
The forms of the both potentials and electric fields are similar in between, but absolute values for  the Moliere potential are
 larger approximately on 10 \%  than for experimental one. So, the potential bariers are equal to 23.36 eV  and 21.38 eV
for Moliere and experimental approximations, correspondingly (see Fig. 2).   
\subsection{Calculation of volume reflection angle} 
Fig. 3  illustrates the calculated volume reflection angle as a function of the transversal energy for several values of the
radius. As expected,  this function is periodic in  accordance with Eq.(16). However, we see
that $\alpha(E)$ is a discontinuous function.  The purpose is obvious: it is a transition by bound  from one potential cell
to another.  It is significant that function $\alpha(E)$ have  a clear maximum and the value of the maximum  decreases
weakly with the radius decreasing, We find that the appearance of the maximum (or  specifically, when derivative
$d\alpha/dE \ne 0$)
   is connected with the form of 
the atomic potential.  For some model potentials (such as a parabolic or polynomial \cite{MV} (with several numbers of terms)   potentials)
the maximum coincident with the point of discontinuity.
 For mentioned here model potentials is typical incorrect description of the 
electric field (electric field is  a discontinues function near area of atomic location).
Note that with the decreasing radius the  interval of the possible angles increases.
It means that the angle divergence is small at large enough radii.   

Knowledge of the function $\alpha(E)$ allow one to calculate the value $\alpha$ for given quantities $x_0$ and $\theta$.
Fig. 4  illustrate the volume reflection angle as a function of the initial coordinate. 
Here the derivative $d\alpha(x_0)/dx_0$ are also shown.  There are two curves calculated at two different but
very close in between initial angles $\theta$ within angle interval which determined by Eq.(20). This fact shows
the  valuable variation of the $\alpha(x_0)$-function  at small  variations of the initial angle.

\subsection{Initial distributions of particles }
Now we can find the initial distribution over the volume reflection angle and at the fixed initial angle with the help of Eq.(19).
Fig. 5  illustrates our calculations. As  follows from Fig.4  there are three value of angle $\alpha$ where 
the derivative  $d\theta/dx_0$ is equal to zero (at fixed $\theta$). At these angles the distribution function $dN/d\alpha$ tends
to $+\infty$. However, $\int_{\alpha_{-}}^{\alpha_{+}} dN(\alpha) = 1$, where $ \alpha_{-}$ and $\alpha_{+} $
are the minimal and maximal values of $\alpha$. Comparing the two distribution at different values of $\theta$-angle
we see that only maximal value of $\alpha$-angle is the same, but the couples of other angles (at which the derivative are
equal to zero)  are different. This fact is explanation of a disappearance of  the two peaks at the next averaging 
over $\theta$-angle in accordance with Eq.(21). 

Our results of calculation according Eq.(19) we compared with the calculations by the Mote Carlo method of the same 
distributions and got very good agreement in between.

The averaged over initial angles distributions are shown in Fig. 6  for several radii. As expected the peak at
maximum angle is conserved. 
\subsection{Entering angle distributions }
As  was considered  the multiple scattering distorts the initial particle distribution.
For calculations of this factor we can find the distance where particle is scattered according to relation:
$l_c=R\theta$. Decreasing of angle $\theta$ decreases the distance $l_c$.  On the other hand,
the inequality $l_c \gg R\theta_c$ should be fulfilled. Thus, we can find that the  optimal 
angle $\theta$ lies in the angle range $5 10^{-5} - 10 ^{-4}$ radian.  It corresponds to $l_c=0.125 - 0.25 $ cm
for $R=25$ m.  We also should take into account the distance $l_d$ after critical point (see Fig. 1).
Our estimation (see Eqs.(23),(24)) show that the distribution form of the entering beam depends on
the total distance $l_s= l_c+ l_d$ and practically independent of the relation between $l_c$ and $l_d$. 
Fig. 7 illustrates the results of calculations of the entering angle distributions at different radii.
The initial (incoming) distribution we assume the physically narrow.
The results of calculations are partially reflected in the table 1. 
From our calculations one can see that the form of the outcoming
angle distributions is close to Gaussian one.
 
The angle width of the incident beam in CERN experiment is expected 
about some microradians. Then assuming  its Gaussian form it is easy
to find the resulting angle distribution after the bent single crystal. 
At the present time the precise Monte Carlo calculation of the angle distributions for
conditions of the future experiment in CERN were made \cite{TA}.  However, 
we can say that our analytical description is consistent with these  results
and presents the additional information.

\subsection{Volume reflection of negatively charged particles}
We carried out some calculations of  parameters of volume reflection effect
 for negatively charged particles at energy 400 GeV. In particular, we
calculate the angle of volume reflection as a function of the transversal energy.
For large enough radii ( $\approx 25 m$), the result is similar as in Fig. 3 but
the value of maximal angle is less in $\approx 1.8$ times.

It should be noted, that  one can expect   some peculiarities  in the  volume reflection
at small enough radii of bending for both sorts of charged particles.  
Really at very small radii the effect of volume reflection should be disappear
due to dominant value of the linear term in effective potential. Mathematically,
the next condition should be satisfied: $\delta E \gg U_0$, where $U_0$ is
the potential barrier of the usual one dimensional potential.
For 400 GeV protons (antiprotons) the equality $\delta E = U_0$ takes place
at $R\approx 3.5$ m. We remind that through the paper only the (011) plane of
silicon single crystal is considered. 

At the bending radii $ \sim 3.5$ and less the angle of reflection (for both sorts
of charged particles) may be of the different signs, or, in other words,
the particles may be scattered in the opposite sides. 
Fig. 8 shows the angle of volume reflection as a function transversal energy
for $R=1$ m and negatively charged particles. One can see that the particles scattered mainly through relatively
small angles, but there is the narrow range of a strong scattering through
angles till $7 10^{-5}$ radian. Partially, we can explain this fact by the specific
form of the effective potential (see Fig.8b).  We can also explain the discontinuity
of $\alpha(E)$-function (see Fig. 8a) by this reason.

Now using Eqs.(23) and (24)  we find the angle distribution after single crystal for
physically narrow initial beam. For calculations we select the thickness
of the single crystal $\approx 0.01$ cm. Fig. 9 illustrates the
results of our calculations. From here we get that 
$\approx 10, \,3, \, 1 $ percents of the beam scattered through angle
more than $10^{-5},\, 2 10^{-5},\, 3 10^{-5}$ radians, correspondingly.   
\section{Conclusions} 
Considered in paper analytical description of the process of volume reflection
of charged particles in bent single crystals allow one to find
necessary characteristic of scattered beam.
We think that suggested method may be useful at different investigations
at interactions  particles with single crystals.
This method may be used as additional to Monte Carlo calculations.
Further development of our description is needed:

1) the process of the volume capture of particles in channeling regime
should be included in consideration with aim of the full description
of the process;

2) the extension of the method on single crystals with the
variable radius is very desirable.

We suppose that the first point  may be easy enough realized 
because the main cause of the phenomenon is well-known.
We believe that the second point may be also solved if
take into account that near the critical point the confined periodicity
of the volume reflection angle  takes place in some small volume of
the bent single crystal.

\section{Acknowledgements}
Author would like to thank all the colleagues-participants of the 
CERN channeling experiment for attention and fruitful discussions.

This work was partially  supported
by the grant INTAS-CERN 05-103-7525.


\vspace{70 mm}
\newpage
\begin{figure} 
\begin{center}
\parbox[c]{14.5cm}{\epsfig{file=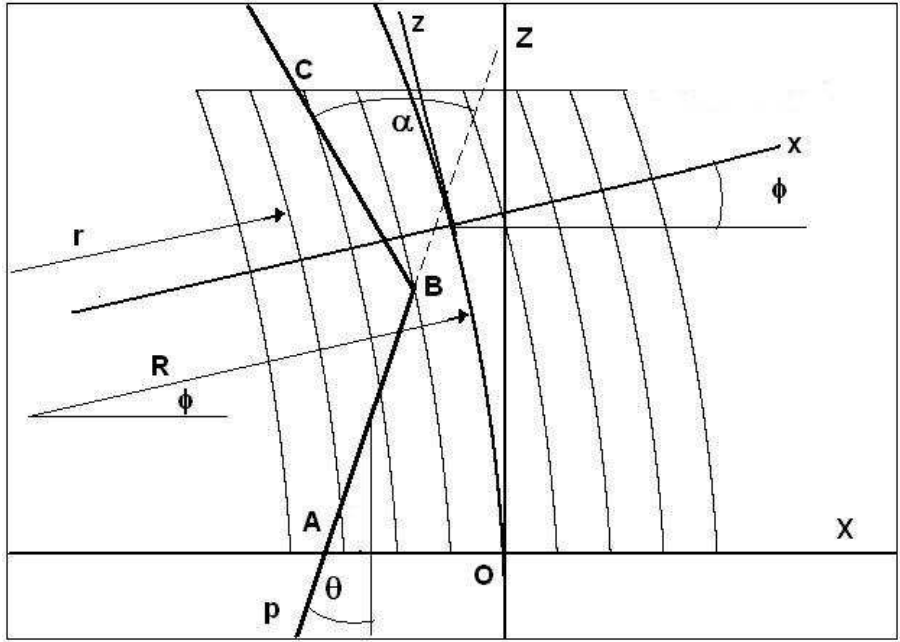,width=12cm}}
\parbox[c]{15cm}{\caption{
 Scheme of the volume reflection of the proton beam. $XYZ$ is the Cartesian coordinate system at the
entrance in single crystal, $xyz$ is the local Cartesian coordinate system connected with the current location 
of the particle. $Y$-axis is directed normally to plane of figure. $\theta$ and $\alpha$ are the initial and  volume
reflection angles.
              }}enough 
\end{center}
\end{figure}
\newpage
\begin{figure} 
\begin{center}
\parbox[c]{14.5cm}{\epsfig{file=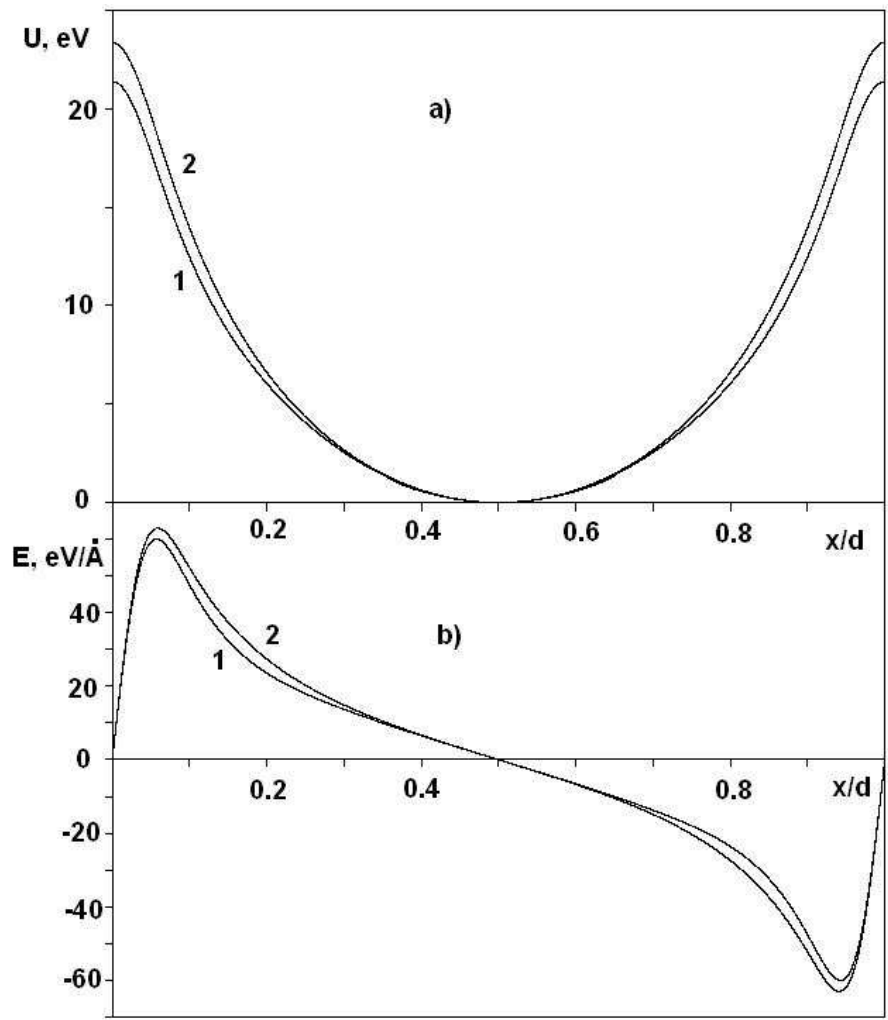,width=12cm}}
\parbox[c]{15cm}{\caption{
 One dimensional potential (a) and electric field (b) in the (110) plane of the silicon single crystal (at room temperature).
as  functions of the relative coordinate $x/d$, where $d=1.92$ angstroms is the interplanar distance.
The curves 1 and 2 correspond to calculations based on the experimental and Moliere atomic form factors.
              }}
\end{center}
\end{figure}
\newpage
\begin{figure} 
\begin{center}
\parbox[c]{14.5cm}{\epsfig{file=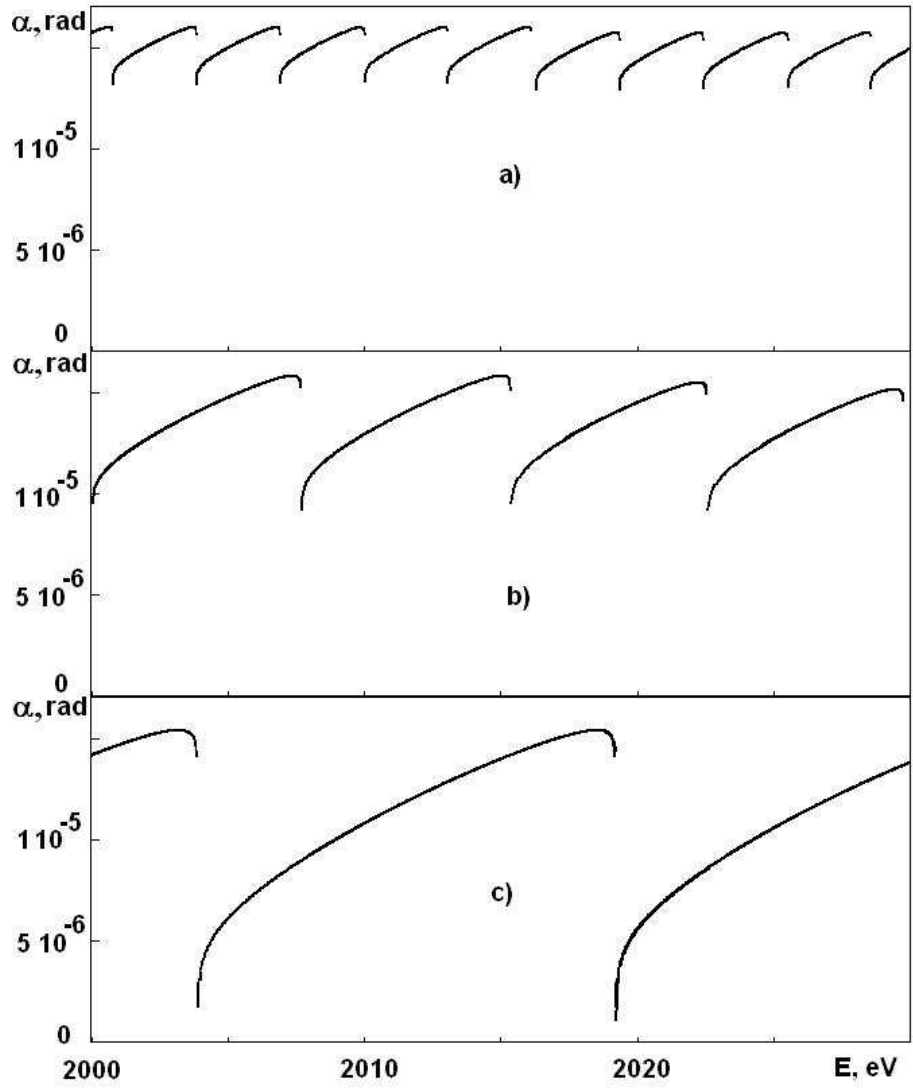,width=12cm}}
\parbox[c]{15cm}{\caption{
 Angle of volume reflection as a function of the transversal energy for
the values of radius equal to 25 m (a), 10 m (b) and 5 m (c), respectively. 
              }}
\end{center}
\end{figure}
\newpage
\begin{figure} 
\begin{center}
\parbox[c]{14.5cm}{\epsfig{file=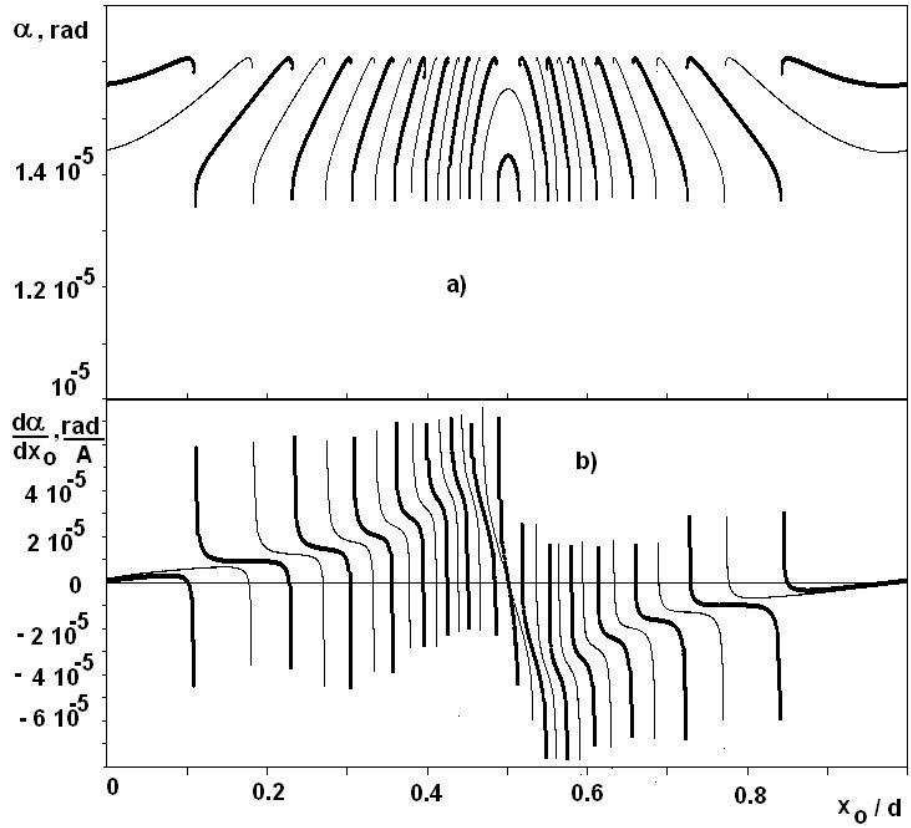,width=12cm}}
\parbox[c]{15cm}{\caption{
 Angle of volume reflection (a) and its derivative (b) as functions of the initial relative transversal
coordinate $x/d$ for radius equal to 25 m.  The thick curves correspond to initial angle $\theta= 10^{-4}$ radian
and thin curves correspond to initial angle $\theta = 10^{-4}+ 5.3 10^{-7}$ radian  
              }}
\end{center}
\end{figure}
\newpage
\begin{figure} 
\begin{center}
\parbox[c]{14.5cm}{\epsfig{file=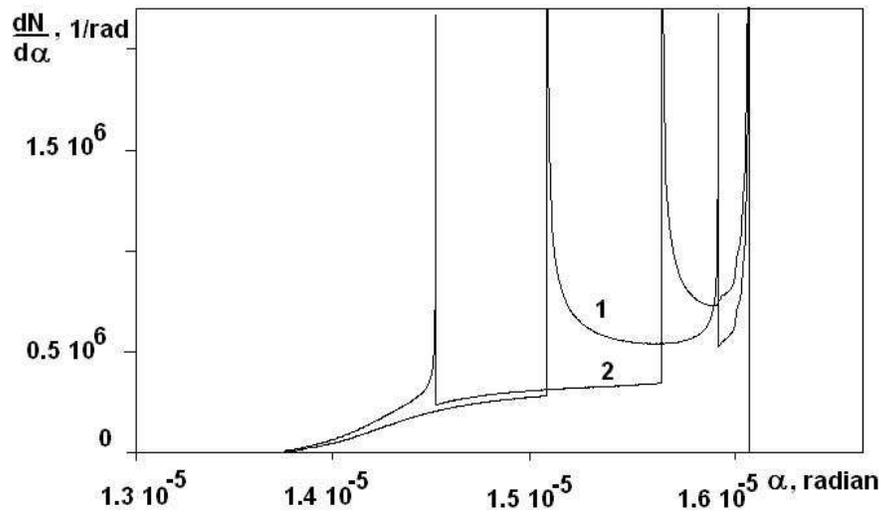,width=12cm}}
\parbox[c]{15cm}{\caption{
 Distribution of particles over the angle of volume reflection. Curves 1 and 2 calculated for the two different initial angles
(with the difference of these angles  $ < \, \delta\theta)$ nearby $\theta=0.0001$ radian and for $R=25$ m. 
              }}
\end{center}
\end{figure}
\vspace*{100 mm}
\newpage
\begin{figure} 
\begin{center}
\parbox[c]{14.5cm}{\epsfig{file=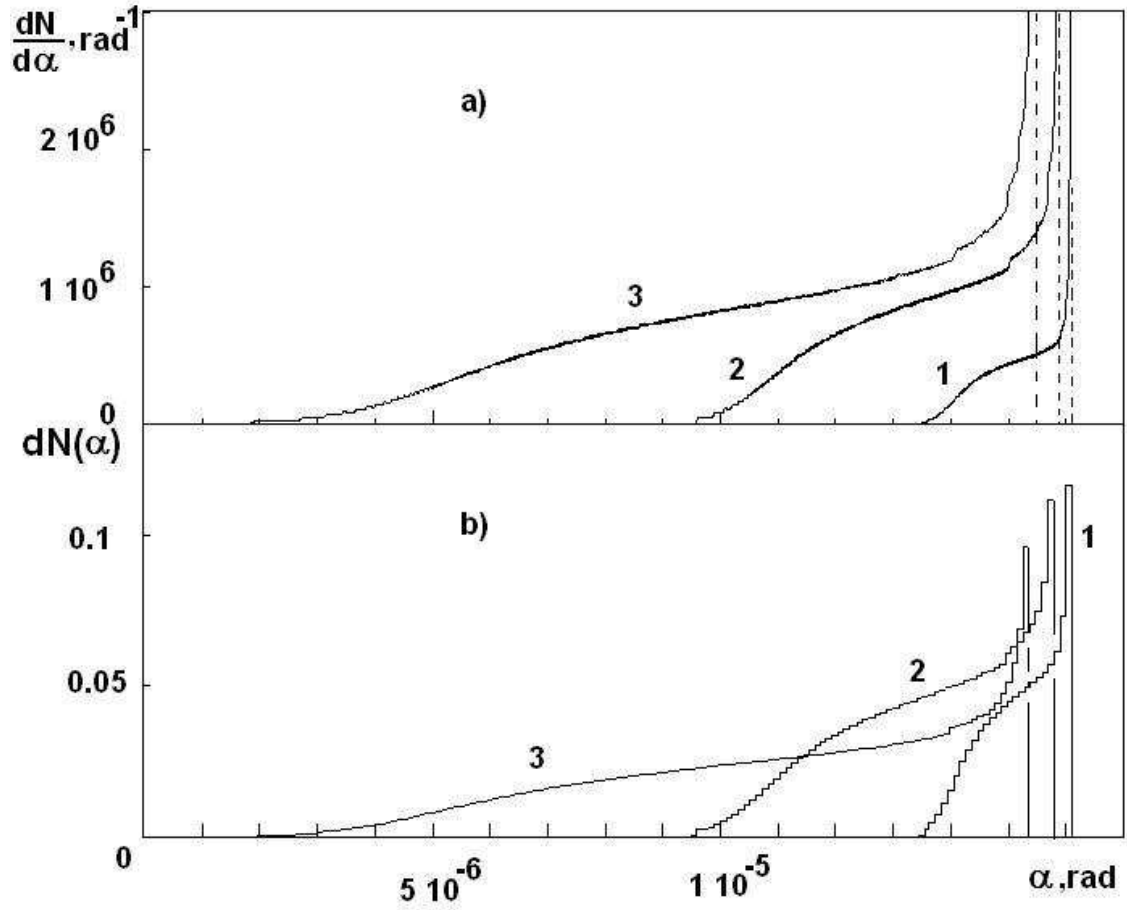,width=15cm}}
\parbox[c]{15cm}{\caption{
 Averaged over period of initial angles nearby $\theta=10^{-4}$ radian,  distributions of particles over the angle of volume reflection (a) .
For visualization,  the step distributions (b) were obtained by integration of the distributions (a) in angle intervals equal to $10^{-7}$ radian. 
Curves 1, 2 and 3 correspond to values of radius equal to 25, 15 and 10 m, respectively.
For visualization,  the y-coordinate for curves 2 and 3 are increased in 5 and 10 times for (a) and in 5 and 12 times for (b) cases.
              }}
\end{center}
\end{figure}
\newpage
\begin{figure} 
\begin{center}
\parbox[c]{14.5cm}{\epsfig{file=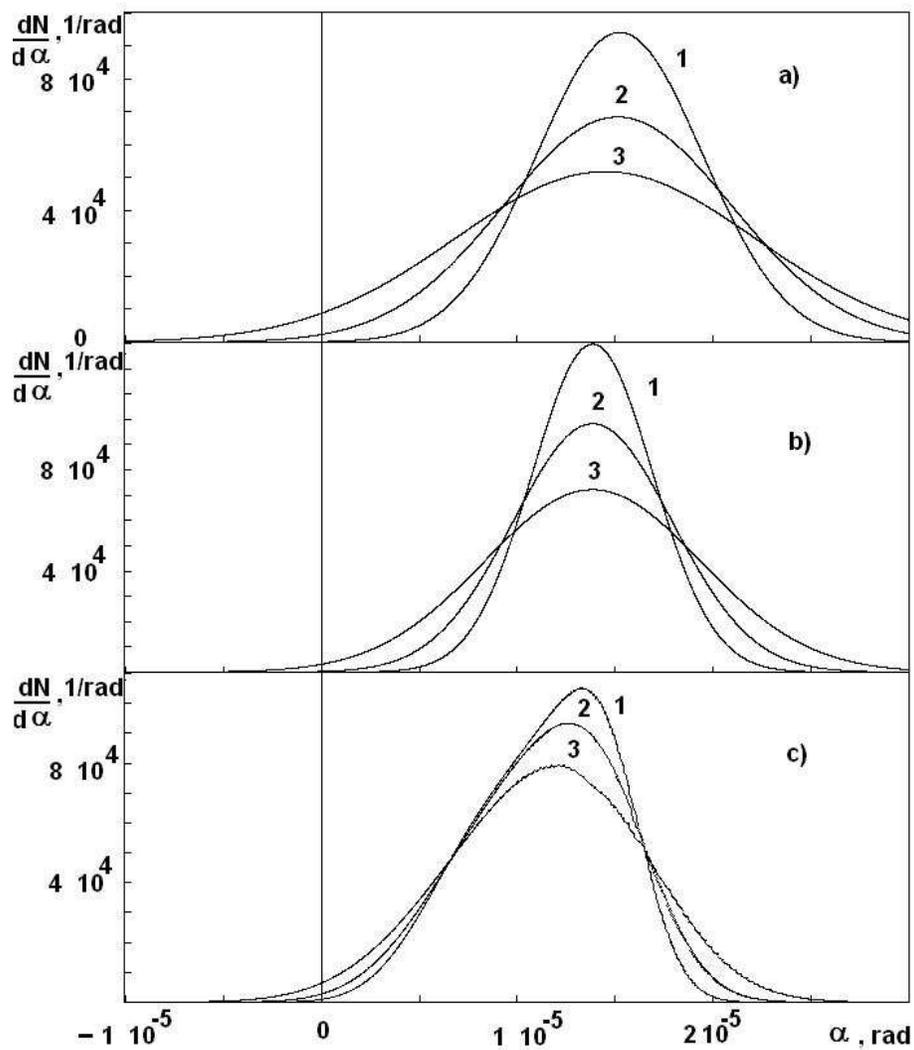,width=12cm}}
\parbox[c]{15cm}{\caption{
 Distributions of particles after volume reflection and multiple scattering (at the output of single crystal) as functions (
of the deflection angle. Zero angle corresponds to the initial direction of the beam.  Calculation was made for
different radii: 25 m (a), 10 m (b) and 5 m (c). For additional information see text and table. 
              }}
\end{center}
\end{figure}
\newpage
\begin{figure} 
\begin{center}
\parbox[c]{14.5cm}{\epsfig{file=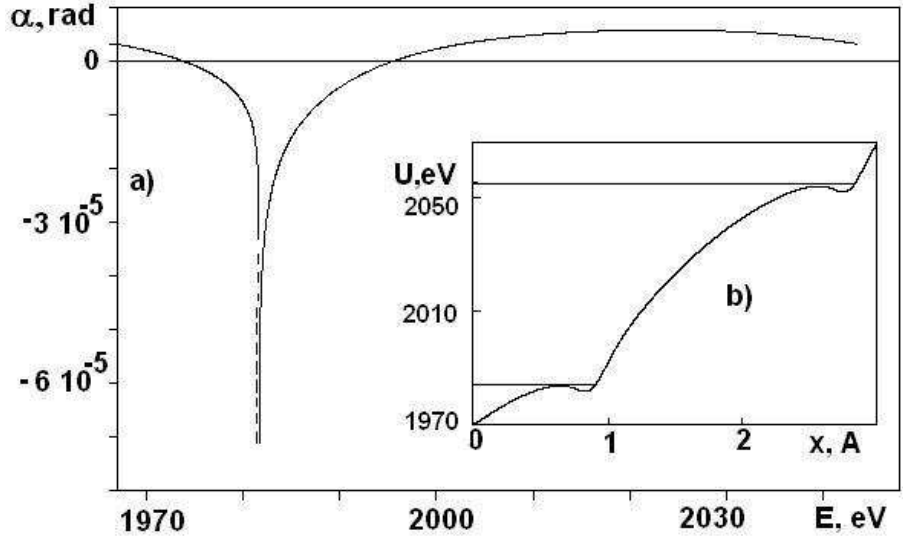,width=12cm}}
\parbox[c]{15cm}{\caption{
 One period of energy dependence of the volume reflection angle (a) and effective potential (b)  for
negative particles and for $R= 1$ m.  The dotted line (in (a)) shows the discontinuity of a function.
 The horizontal lines (in (b)) correspond to energies at which the discontinuity takes place.  
              }}
\end{center}
\end{figure}
\vspace{100  mm}
\newpage
\begin{figure} 
\begin{center}
\parbox[c]{14.5cm}{\epsfig{file=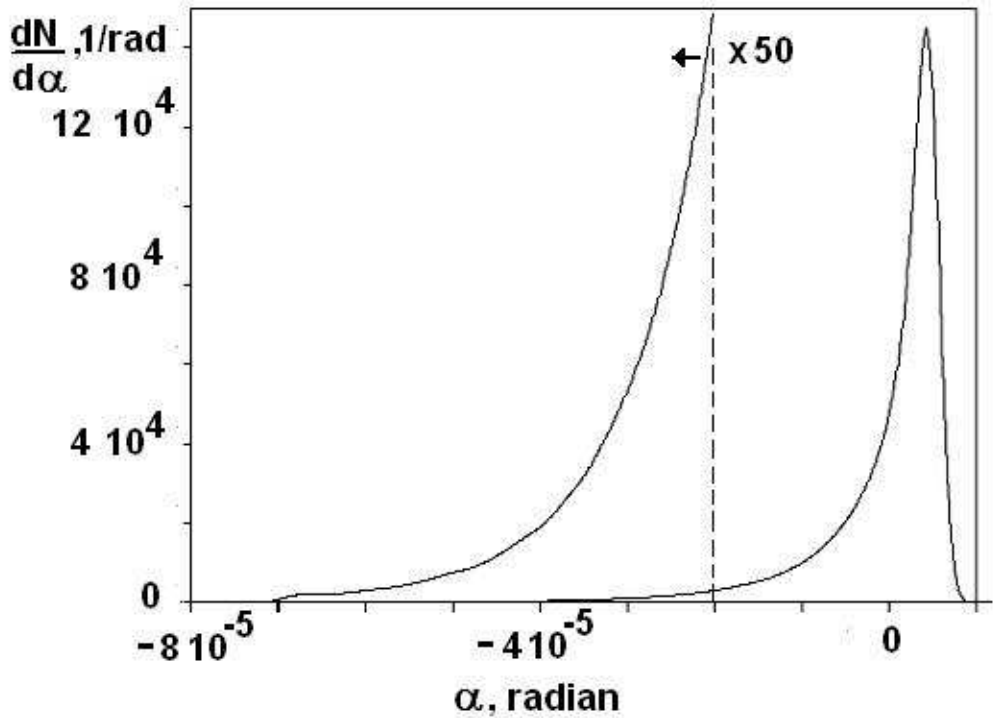,width=15cm}}
\parbox[c]{15cm}{\caption{
 Angle.distribution of the negative particles after bent single crystal (R= 1m).
The left part of the distribution are also shown with increasing of scale in 50 times.
The thickness of  the single crystal is equal to 0.01 cm.
              }}
\end{center}
\end{figure}
\end{document}